# A PUF-Based Security Framework for Fault and Intrusion Detection

Ahmed Oun[1], Rishabh Das[2], Clay Hess[1],
Aakriti Barat[1], Savas Kaya[1]

[1]School of Electrical Engineering and Computer Science,
Russ College of Engineering and Technology, Ohio University, OH, USA
Emails: {oun, ch559521, ab613319, kaya}@ohio.edu

[2]J. Warren McClure School of Emerging Communication Technologies,
Scripps College of Communications, Ohio University, OH, USA
Email: rishabh.das@ohio.edu

*Abstract*—Industrial Control Systems (ICS) rely on sensor feedback to keep safety-critical processes within operational limits. This research presents a hardware-root-of-trust that embeds a Physically Unclonable Function (PUF) at the measurement layer to authenticate sensor readings. The architecture combines voltage fingerprinting with a temporal authentication that integrates with standard industrial control system architecture. The research prototypes the PUF integration on a hardware-in-the-loop (HIL) water tank testbed using a Simulink-based PUF emulator. The system maintains 99.97% accuracy over a 5.18-hour period of normal operation and flags all injected anomalies, including spike faults, hard-over faults, and hardware trojan scenarios that push the system over to an unsafe operational state. The proposed architecture provides a process-aware, vendor-agnostic approach that can integrate with legacy plants to detect sensor signal degradation or sophisticated supply chain attacks.

*Index Terms*—Hardware Security, PUF, ICS, Trojan.

## I. INTRODUCTION

Industrial control systems (ICS) like water treatment plants, power generation stations, chemical refineries, and transportation networks manage the infrastructure necessary for the socio-economic needs of a nation [1]. ICS systems rely on interdependent feedback loops, real-time sensor measurements, and actuation for safe operation. The sensor measurement provides visibility into the state of the industrial systems and allows the controller to manage the operations. An attack on the measurement layers, resulting in a compromised sensor reading, can make the control logic act on false data, steering the industrial process towards a dangerous state. Measurement layer attacks frequently arise from vulnerabilities within the supply chain. For example, a chemical plant using a level or flow sensor from an unknown manufacturer might have an embedded Trojan [2]. The Trojan can stay dormant for a period of time before activating and reporting spoofed readings. Attacks on the measurement layer impact the operation of the controller, and upstream security tools like next-generation firewalls or network monitoring do not have visibility into the attack. As a result, researchers are investigating methods that can integrate with the industrial process and validate the authenticity of the sensors' readings.

Besides malicious tampering, faults due to electromagnetic interference, vibration, corrosion, and thermal drifts can cause similar effects [7]–[9]. Having a process-agnostic true anchor at the sensing layer can detect signal malfunctions. As the demand for secure and efficient authentication rises, traditional methods depending on passwords or stored cryptographic keys face increasing risks from physical attacks and unauthorized access [10]. Physical Unclonable Functions (PUFs) present a hardware-based alternative by leveraging uncontrollable manufacturing variations to create unique, unclonable identifiers for each chip [11]. These variations enable the generation of Challenge-Response Pairs (CRPs), which serve as dynamic, hardware-rooted cryptographic keys, eliminating the need for key storage and significantly enhancing resistance to tampering and key extraction [12]. This research proposes a hardware-based PUF authentication framework that leverages CMOS process variations for secure integration. By embedding security at the hardware level, it reduces the vulnerabilities and enhances protection against attacks, particularly in mission-critical environments like Industrial Control Systems (ICS). The approach also incorporates Total Harmonic Distortion (THD) analysis using an Instrumentation Front Module (IFM) to ensure consistent and reliable device authentication over time. The contributions of this research are:

- A process-aware Physically Unclonable Function based sensor authentication framework that combines analog voltage fingerprints with temporal authentication checks to detect gradual or abrupt deviation in sensor output.
- A hardware-in-the-loop (HIL) testbed architecture to operationalize PUF-based authentication in a water tank control system.
- An empirical evaluation of sensor fault and adversarial scenarios, demonstrating high detection accuracy.

The rest of the paper proceeds as follows. Section II outlines the architecture of the PUF, the enrollment process, and Section III outlines the PLC integration and results. Section

IV concludes the work and describes the future research areas.

## II. METHODOLOGY

This section outlines a methodology that automatically baselines the characteristics of industrial sensors using a Physically Unclonable Function (PUF). The PUF integrates with the industrial control system and is the trust anchor. The PUF enables the real-time validation of incoming sensors' readings and identifies changes in voltage characteristics due to operational issues like electrical faults or cyber threats like hardware-based spoofing. The proposed design integrates with traditional industrial control system architecture and requires no architectural change.

The following subsections discuss the integration of the PUF with the industrial control system architecture, the architecture of the PUF, the enrollment phase, and outline the overall workflows of the sensor value validation.

### A. Overall workflow and ICS architecture

An Industrial Control System has five major components: (1) the physical system which includes the sensors and actuators to monitor and control process variables, (2) the cyber-physical link which provides a transport medium for the signals from the sensors and actuators to the controller, (3) a programmable logic controller (PLC) monitors the sensor readings and takes action based on a predefined logic, (4) the industrial network connects the controller with other subsystems, and (5) the Human Machine Interface (HMI) is the graphical user interface that allows operators to interact with the control system. In the proposed architecture, the PUF uses the TCP-MODBUS industrial network protocol to communicate with the programmable logic controller (PLC). The PUF reads the value of the sensor's voltages from the PLC's Input register and initiates the authentication process. The authentication process returns a "3" if the value of the sensor aligns with the learned baseline or a "1" if the value does not match the baseline. The PUF writes the response to the PLC's memory. By analyzing the response, an operator monitoring the PLC via a human-machine interface can identify discrepancies in the sensor readings.

### B. PUF Architecture

The proposed Physically Unclonable Function (PUF) architecture leverages inherent manufacturing variations in CMOS inverter transitions as its entropy source, as shown in Figure 1 [13]. To generate a response bitstream, an array of inverters is utilized. For a given challenge input, the transition regions of these inverters are swept to extract the corresponding total harmonic distortion (THD). The consequent THD value is then compared to a precomputed mean value stored in a lookup table (LUT), which is indexed by the same challenge. This LUT comprises 256 entries, each representing the average THD response across multiple samples for a given challenge. The outcome of this comparison yields a single bit of the final PUF response.

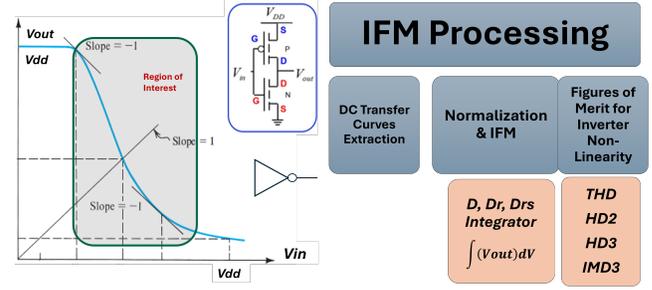

Fig. 1: The proposed approach employs the Integral Function Method to extract both linearity and distortion characteristics from the quasistatic curves of CMOS inverters.

In order to obfuscate the PUF functionality and enhance security, the challenge seeds a pseudo-random number generator (PRNG). The PRNG dynamically determines distinct sweep regions for each inverter, ensuring variability in the response generation process. The implementation of the PUF uses an 8-bit challenge space and operates over 18 inverter curves, thereby producing an 18-bit response vector. The inverter characteristics are modeled as pre-recorded data arrays stored in MATLAB files and are used as input to the PUF module. In a hardware realization, only the challenge would be provided as input to the PUF, and the inverter responses would be generated physically.

### C. Verification Module Architecture

The verification module is responsible for authenticating system behavior by using four primary inputs: the current tank level, the PUF response, the previous output, and an enrollment flag, as shown in Figure 2.

*1) PUF Verification:* The PUF Verification operates based on a predefined set of verification thresholds and an associated tolerance window. The verification thresholds consist of a series of discrete tank level values used for authentication. In our implementation, these values range from 0 to 300 in increments of 20 (i.e., 0, 20, 40, ..., 280, 300). The tolerance window defines the acceptable deviation from each threshold within which a measurement is considered valid for verification or enrollment. This window is set to ±2 units. For example, consider an actual tank level of 98. In that case, it falls within the [98, 102] window of the 100 threshold, and the verification module proceeds with enrollment or verification based on the state of the enrollment flag. Conversely, if the tank level is 85, it lies outside the tolerance window of any defined threshold. In such cases, the verification module retains its previous output and performs no new authentication operation. This mechanism ensures that sensor readings are only processed when they are sufficiently close to known, trusted operating points, reducing the risk of noise-triggered false authentication.

*2) Temporal Verification:* The temporal verification operates by maintaining a queue of predefined length that tracks the most recent sensor readings. Temporal verification runs regardless of the tank level. The verification works by tracking the difference between the maximum and minimum elements of the queue. During the enrollment phase, the temporal verification will record the maximum observed difference in the queue. During the authentication phase, the maximum difference within the queue is compared to the enrollment value. If the difference exceeds the enrollment value, the temporal verification bit of the response is set to 0 until the temporal reset flag is set high.

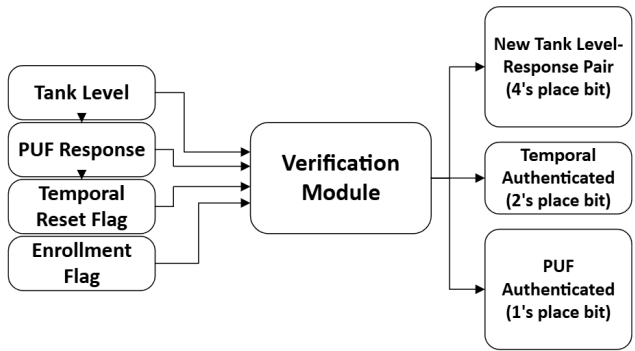

Fig. 2: Verification Module block diagram.

### D. Enrollment Phase

Once the enrollment flag is asserted (i.e., set high), the verification module enters enrollment mode, during which it is assumed that the sensor under observation is functioning correctly and without faults. In this mode, the module records pairs of tank level values and their corresponding PUF responses to memory for future authentication. To ensure data relevance and integrity, only tank-level values that fall within the defined tolerance window of one of the predefined verification thresholds are considered for storage. Furthermore, a tank level-PUF response pair is recorded only if it has not been previously stored, preventing duplication and conserving memory resources. Also, during this phase, the verification module records a threshold. The module provides feedback on enrollment activity through the 4's place bit in its output:

- A value of "1" indicates that a new unique tank level-PUF response pair has been successfully recorded.
- A value of "0" indicates that the observed pair has already been stored during a previous enrollment cycle.

The system must remain in the enrollment phase for a sufficient duration to observe and record the full range of expected tank level and PUF response combinations. An insufficient enrollment phase can lead to incomplete reference data, thereby increasing the possibility of false positives during subsequent authentication operations.

### E. Signal Authentication Phase (Challenge-Response)

When the enrollment flag is low, the verification module operates in authentication mode, performing real-time validation of the sensor signal using previously recorded data from the enrollment phase. During this phase:

*1) PUF Authentication Phase:*
- The module first checks if the current tank level falls within the predefined tolerance window (i.e., within ±2 units in our implementation) of any registered authentication threshold (e.g., 0, 20, ..., 300).
- If the tank level meets this condition, the module retrieves the current PUF response and attempts to locate the corresponding tank level–PUF response pair in memory.

The authentication decision logic is as follows:
- If the pair is found in memory, the module sets the least significant bit to "1," indicating that the sensor reading is authenticated and matches a previously enrolled valid state.
- If the pair is not found in memory, the module sets the least significant bit to "0," signaling a sensor fault, anomaly, or potential attack.

If the current tank level does not fall within any authentication window, the verification module retains its previous 1's place bit, avoiding an unnecessary authentication attempt. This lightweight challenge-response verification ensures that only expected and previously seen sensor behaviors are accepted as valid, thereby providing a mechanism for detecting spoofed, faulty, or malicious sensor data in real-time.

*2) Temporal Authentication Phase:* During the authentication phase, the verification module continues to maintain a queue of sensor data and calculate the difference between the maximum and minimum values in the queue - as it did in the enrollment phase. The continually calculated difference is compared with the maximum difference from the enrollment phase. The temporal output bit (the 2's place bit) will stay high so long as the calculated difference falls below the maximum difference from the enrollment phase. Should the calculated difference exceed the maximum difference from the enrollment phase, the temporal authentication fails and the module outputs a "0" for the temporal output bit; until the temporal reset bit goes high, the temporal output bit will stay low, even if the calculated difference falls below the maximum difference from the enrollment phase.

## III. RESULTS

### A. Experimental Setup

The experimental setup uses a hardware-in-the-loop water tank system consisting of four major components: a simulated water tank system, an industrial network using TCP-MODBUS, a hardware PLC, and a simulated PUF for validating sensor characteristics. The physical system uses a 3D simulation platform called Factory IO. The Factory IO simulates the physics of the water tanks, sensors, and actuators, and communicates the values to the PLC. This research uses an Open-source PLC called OpenPLC [3] with structured text logic to support the control of the testbed. The OpenPLC runs on a networked Raspberry Pi 4 compute board. A Python script connects to the PLC using TCP-Modbus and simulates human-machine interface (HMI) actions. Articles [4], [5], and [6] outline the architecture of the virtual industrial control system.

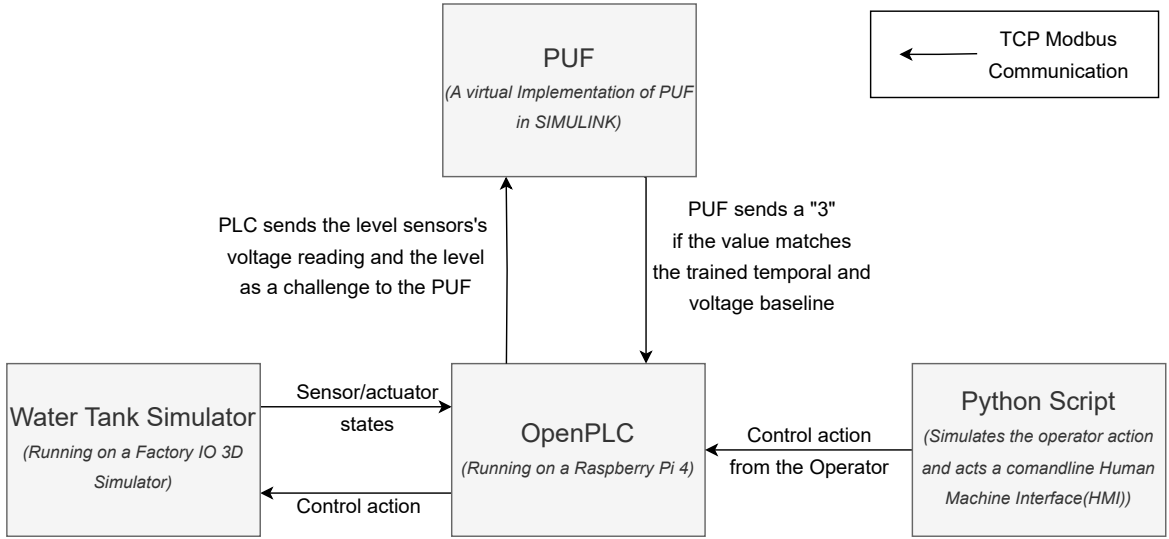

Fig. 3: Experimental setup showing the integration of PUF with the water tank industrial control system

The OpenPLC, Python script, and the Factory IO provide the essential building blocks of a complete industrial control system. The PUF communicates with the PLC to learn and then validates the sensor readings.

The water tank testbed offers two modes of operation: a manual mode and an auto mode. In manual mode, the PLC does not control the operations, and the system is driven by operator interactions. In auto mode, the PLC uses an ON/OFF control and maintains the water between a low and a high setpoint. The architecture of the water tank testbed is shown in Figure 3.

### B. Normal Plant Operation

After enrolling the PUF on the water tank system, we simulate normal operation. During normal operation, the water tank testbed is set to auto mode, and the PLC maintains the level between the high and low setpoints. We use a Python script to simulate operator behavior. The script uses the ModbusTcpClient library and writes three key values to the specified holding registers: a low setpoint, a high setpoint, and a valve position. The valve position is randomly set to either ON or OFF, and the low and high setpoints are randomized. After writing the values, the script sleeps for a period of time and then repeats the process.

During normal operation, the PUF should flag all values as normal with a response of "3". "3" highlights a successful temporal and PUF authentication. This experiment collects the PUF response over 5.18 hours of normal operation (286,281 samples) and analyzes the responses for each time step. The results show a 99.97% accuracy with 286,209 samples correctly classified as normal. Seventy-two samples show a "2" response, indicating successful temporal alignment but deviation from the voltage-level baseline. It is important to note that the PUF captures the temporal trend perfectly and does not report any anomaly in the temporal trend of the level sensor during the normal operation of the water tank. A longer enrollment period might reduce such false-positive deviation in the voltage-level baseline. The high accuracy shows the PUF's success in baselining the water tank process.

Table I outlines the accuracy of the PUF over 30-minute windows. The accuracy stays stable, and 6 out of 11 windows show perfect response. During the 5.18 hours of normal operation, the PUF shows a false-positive rate of 0.025%.

TABLE I: Accuracy of PUF by 30-minute period

| Period | Time | Total Samples | Correct Classification | Accuracy |
|---|---|---|---|---|
| Period 1 | 0:00 | 26768 | 26768 | 100.00% |
| Period 2 | 0:30 | 26890 | 26890 | 100.00% |
| Period 3 | 1:00 | 26784 | 26771 | 99.95% |
| Period 4 | 1:30 | 26935 | 26935 | 100.00% |
| Period 5 | 2:00 | 26800 | 26785 | 99.94% |
| Period 6 | 2:30 | 26892 | 26877 | 99.94% |
| Period 7 | 3:00 | 26909 | 26896 | 99.95% |
| Period 8 | 3:30 | 28537 | 28537 | 100.00% |
| Period 9 | 4:00 | 29624 | 29608 | 99.95% |
| Period 10 | 4:30 | 29624 | 29624 | 100.00% |
| Period 11 | 5:00 | 10518 | 10518 | 100.00% |

### C. Sensor Fault Detection

Sensors can withstand various environmental conditions, but aging, wear-and-tear, physical damage, or electromagnetic interference can generate incorrect responses [7], [8]. Prompt detection of sensor faults provides additional visibility into the state of the control system. The operator can take countermeasures to stabilize the control loop [9]. This section simulates two sensor fault scenarios and evaluates the PUF's response under each condition.

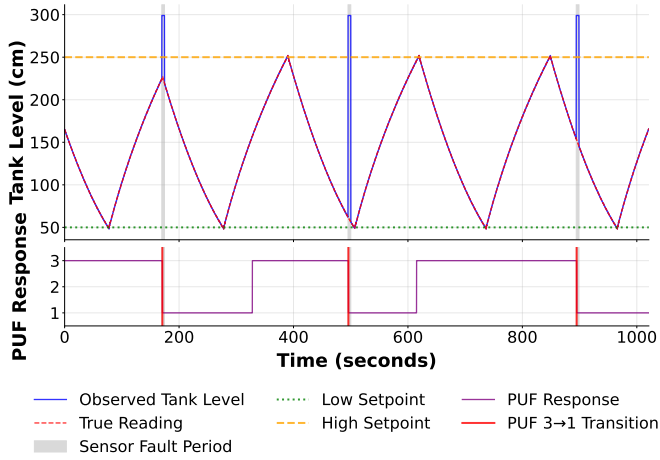

Fig. 4: Response of the PUF during level sensor spike fault

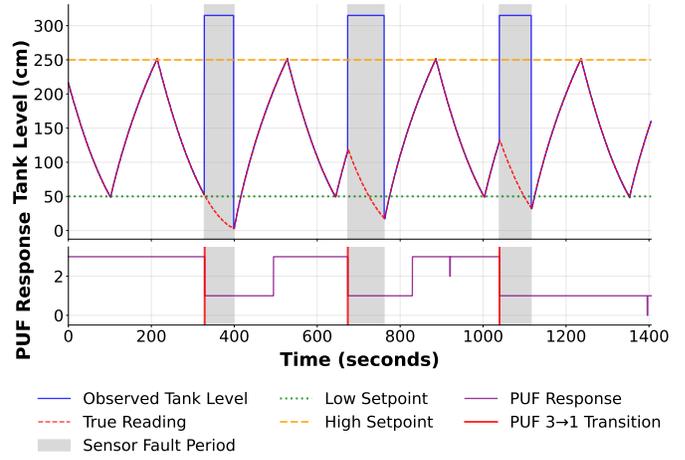

Fig. 5: Response of the PUF during level sensor positive hard-over faults

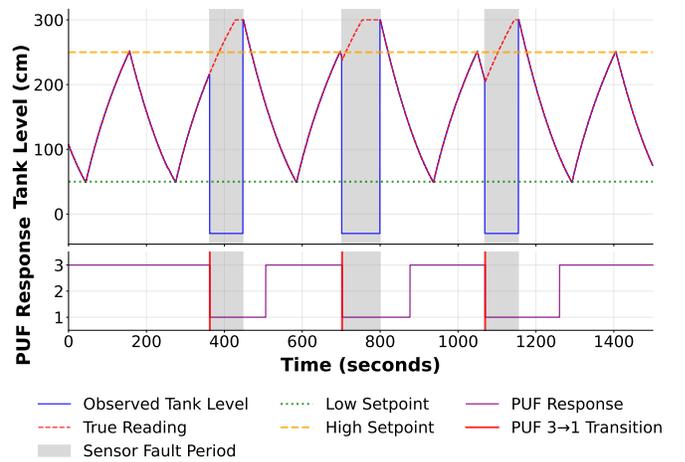

Fig. 6: Response of the PUF during level sensor negative hard-over faults

*a) Spike Fault:* A spike fault generates a high-amplitude impulse signal that can go beyond the expected range of normal operation. Spike faults can happen due to transient electromagnetic interference, momentary short circuits in cabling, or static discharge [9]. In a closed-loop control system like the water tank system, this can provoke an aggressive corrective action. These aggressive actions include abrupt valve closure or toggling of the pump control, which can destabilize and cause safety interlocks. In this experiment, we simulate a spike fault on the level sensor. A data collector inside the PLC analyzes the response of the PUF and evaluates the detection performance.

Figure 4 shows the PUF's response to three spike faults injected into the level-sensor signal. The water-tank process operates in automatic mode with high and low setpoints fixed at 250 and 50, respectively. Each fault introduces a high-amplitude impulse lasting 4.59 s, 4.89 s, and 4.94 s. When a spike occurs, the PUF flags a temporal authentication failure, transitioning from normal (3) to alert (1). After verifying the alert, the operator manually resets the PUF to resume monitoring.

*b) Hard-Over (or Abrupt) Fault:* An abrupt step shift drives the sensor output to its upper (or lower) limit, often overshooting and locking the reading at boundary values [7], [8]. Open circuit wiring or a failed amplifier can cause a hard-over fault. Since hard-over faults can cause extreme changes in sensor values, the controller can initiate an emergency shutdown logic or make abrupt changes to the control system.

In this experiment, we simulate positive and negative hard-over faults on the level sensor. A positive hard-over drives the sensor voltage into high-rail saturation, causing the reported level to exceed the instrument's upper limit. We injected three positive hard-over events lasting 70.49 s, 87.99 s, and 77.16 s. The controller does not receive trustworthy readings during the faults, and the PLC fails to maintain the water level above the low setpoints. Figure 5 shows the true and faulty sensor readings. The PUF detects a temporal authentication failure during fault and transitions from normal (3) to alert (1).

A negative hard-over fault collapses the level-sensor voltage to its lower rail, causing the sensor to report a level of zero. Lacking reliable feedback, the controller permits the water level to exceed the high setpoint, which results in an overflow. In this study, we introduced three negative hard-over events lasting 59.51 s, 59.32 s, and 57.87 s. During each event, the PUF identifies a temporal authentication failure and shifts from its nominal state 3 to its alert state 1. Figure 6 shows the response of the PUF during the negative hard-over faults.

### D. Hardware Trojan Scenario

In the hardware tampering attack, we simulate a hardware trojan scenario introduced through a compromised manufacturing supply chain. The hardware trojan is inside the sensor and stays dormant for the first hour of operation. After an hour, the Trojan auto triggers and reports spoofed readings to the PLC. The trojan subtracts a constant value from the current reading. The PLC controls the water tank based on the spoofed reading, eventually causing a tank overflow scenario. A data collector inside the PLC records the sensor reading and tracks the response of the PUF during attack scenarios.

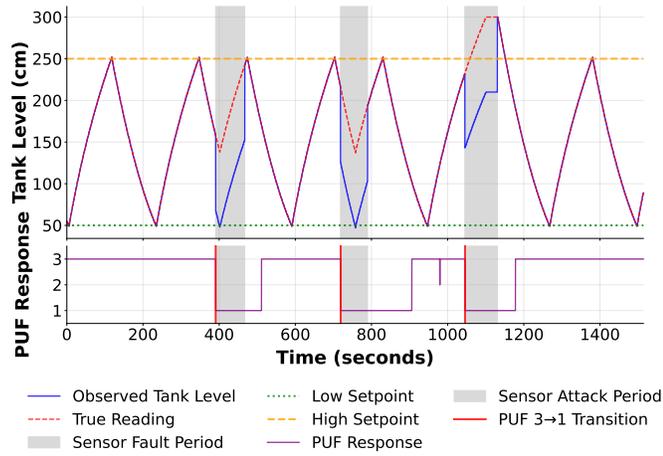

Fig. 7: Response of the PUF during hardware trojan attack

Figure 7 illustrates three trojan activations for periods of 76.31 s, 71.24 s, and 85.07 s, respectively. During the third activation, the actual water level surpasses the high set-point and triggers an overflow. In the first two activations, the controller fails to confine the process within the intended 50–250 level band and drives the tank into an unintended operating range. Throughout all three activations, the PUF identifies each temporal authentication failure and raises an attack alert.

## IV. Conclusion and Future work

This work introduces a hardware root-of-trust that embeds a CMOS-based Physically Unclonable Function at the measurement layer of a HIL water tank system. The PUF reads the level sensor from the programmable logic controller's (PLC) holding registers, authenticates the voltage signature, and temporal trends. The PUF relays the verification result to the PLC's holding register. An HMI can read the PUF response and identify sensor faults or tampering. During a 5.18-hour normal operation, the system shows a 0.025% false positive rate (72 false positives out of 286,281 sensor readings) on the voltage signature authentication and zero false positives on the temporal authentication. The system also flags spikes and hard-over sensor faults.

Future work can migrate the virtual PUF model to a Field Programmable Gate Array (FPGA) that performs a real-time signal validation for every PLC cycle. Researchers can benchmark the approach on different industrial control system domains and test the accuracy in a multi-sensor environment with interdependent controllers. The PUF can also integrate with a trusted execution environment, creating a computing cage for validating sensor response and control operations.